\documentclass[11pt]{article}

\usepackage{color,amsthm,amsmath,natbib,algorithm,algpseudocode}
\usepackage{parskip}
\usepackage{amsfonts}
\usepackage{enumerate}
\usepackage{graphicx}
\usepackage{rotating}
\usepackage{natbib}
\usepackage{hyperref}
\newcommand{\argmin}{\mathop{\arg \min}\limits}

\newcommand{\calN}{\mathcal{N}}

\newcommand{\X}{\mathbf{X}}
\newcommand{\Y}{\mathbf{Y}}

\begin{document}
\title{Network-based multivariate gene-set testing}

\author{Nicolas St\"adler\\
\small Netherlands Cancer Institute\\[-0.8ex]
\small \small  Amsterdam, Netherlands.\\
\small \texttt{n.stadler@nki.nl}\\
\and
Sach Mukherjee\\
\small Netherlands Cancer Institute\\[-0.8ex]
\small \small  Amsterdam, Netherlands.\\
\small \texttt{s.mukherjee@nki.nl}\\
}

 %\footnote{Seminar f\"ur Statistik, ETH Z\"urich, CH-8092 Z\"urich,
 %     Switzerland. staedler@stat.math.ethz.ch, buhlmann@stat.math.ethz.ch}

%generates the title
\date{}
\maketitle

\begin{abstract}

%\textbf{Motivation:} 
The identification of predefined groups of genes (``gene-sets")
which are differentially expressed between two conditions (``gene-set  analysis", or GSA)
is a very
popular 
analysis in bioinformatics.
% and is known under
%the name Gene-set analysis (\cite{subramanian2005,irizarry2009}). 
GSA incorporates biological knowledge by
aggregating over genes that are believed to be functionally related.
%has several advantages
%over gene-by-gene differential expression analysis: Gene-sets obtained
%using biological knowledge 
This can enhance statistical power over analyses that consider only one gene at a time.
However, currently available GSA approaches are all based on univariate two-sample
comparison of single genes.
This means that they cannot test for differences in covariance 
structure between the two conditions.
Yet interplay between genes is a central aspect of 
biological investigation and it is likely that such interplay may differ between conditions. 
%Such approaches can never
%reveal changes in the correlation pattern of genes between two conditions. 
%Prime examples of gene-sets where 
%An important class of gene-sets are biological pathways where genes
%are thought of interacting with each other in specific ways. 
This paper proposes a novel approach for 
gene-set analysis 
that allows for truly multivariate hypotheses, in particular differences in gene-gene networks
between conditions.
Testing hypotheses concerning networks is challenging due 
the nature of the underlying  estimation problem.
%(dis-)similarity. 
Our starting point is a recent, general approach for
high-dimensional two-sample testing.
We refine the  approach and show how it can be used to perform
multivariate, network-based gene-set testing. We validate the approach in simulated
examples and show results using high-throughput data from several  studies in cancer biology.
%\textbf{Availability:} R-package\\

% We propose novel methodology for testing equality of model
%   parameters between two high-dimensional populations. The technique 
%   is very general and applicable to a wide range of models. The 
%   method is based on sample splitting: the data is split into two
%   parts; on the first part we reduce the dimensionality of the model
%  to a manageable size; on the
%   second part we perform significance testing (p-value calculation)
%   based on a restricted likelihood ratio statistic. Assuming that both
%   populations arise from the same distribution, we show that the
%   restricted likelihood ratio statistic is asymptotically distributed
%   as a weighted sum of chi-squares with weights which can be
%   efficiently estimated from the data. In high-dimensional problems, a single data split can result in a ``p-value lottery". To ameliorate this effect, we iterate the splitting process and 
%   aggregate the resulting p-values. This multi-split approach provides
%   improved p-values. 
%   We illustrate the use of our general 
%   approach in two-sample comparisons of high-dimensional regression models (``differential regression") and 
%   graphical models (``differential network"). In both cases we show results on simulated data as well as real data from recent, high-throughput cancer studies.

\vspace{0.5cm}
\end{abstract}

\newpage
\section{Introduction}
%%%%%%%%%%%%%%%%%%%%%%%%%%%%%%%%%%%%
%%Differential Expression and GSEA%%
%%%%%%%%%%%%%%%%%%%%%%%%%%%%%%%%%%%%
Differential expression analysis \citep{tusher2001,loennstedt2002,smyth_linearmodels2004} is one of the most popular statistical analyses in molecular biology, whether for mRNA (including RNA-seq and microarrays), protein or epigenomic data. 
For each variable (or gene, we use both terms interchangeably throughout but note that methods described apply also to other types of data)
expression levels are compared between 
conditions of interest to obtain a measure of significance for the gene, usually accounting for multiple
comparisons.

\citet{subramanian2005} pointed out a number of drawbacks of classical, single gene differential expression analysis, including 
lack of statistical power and difficulties in interpreting significant genes in a biological
context. They addressed these
concerns by an approach called \emph{gene-set analysis} or \emph{GSA}
which sought to test differential expression not at the level of single genes in isolation,
but rather using (pre-defined) groups of biologically related genes called \emph{gene-sets}.
GSA can provide gains in power by taking advantage of the biological knowledge
encoded in gene-set membership: for example, if several members of a certain gene-set  all show a moderate change between conditions, the gene-set as whole may be significant even if its constituent genes would not be significant on a gene-by-gene basis. Furthermore, by providing results at the level of biologically meaningful sets of genes, GSA can aid in interpretation of the results of differential expression analysis.
GSA 
has become one of the most widely used analyses in bioinformatics.
\citet{irizarry2009} provide a self-contained introduction to GSA aimed at a statistical audience; GSA and extensions thereof are further described  in \citet{subramanian2005,efron_geneset2007,jiang2007}.

GSA focuses on multiple genes taken together. However, existing  GSA approaches \citep{subramanian2005,irizarry2009,efron_geneset2007} are based on \emph{univariate
statistics} comparing the two conditions of interest. 
That is, they aggregate several single-gene comparisons to arrive at a gene-set-level statistic and measure of significance. 
In these procedures, 
 for
each gene $j=1,\ldots,p$ 
a two-sample statistic
 $z_j$ is computed.
The   $z_j$'s for genes belonging to specific gene-sets $A_s, s=1,\ldots,S$ are then combined 
to arrive at aggregate scores $a_s$ at the gene-set-level.
Despite the 
usefulness of these approaches they have a major limitation
in that they are all based on single-gene comparisons 
and are therefore 
 inherently univariate in nature.
In particular, changes in
covariance structure between two
conditions cannot be tested by such approaches. However, 
interplay between molecular variables is a fundamental aspect of biology
and it is likely that in many settings differences in covariance or conditional independence structure between conditions may be relevant.
%As an example, the p53 pathway plays a physiological role in controlling the cellular response to
%different types of stresses (such as DNA damage and hypoxia) but is known to be disrupted in cancer cells.
%Such disruption may be reflected in the 
%
% is disrupted in cancer cells. % which then can result in tumor progression. 
%Like for the p53 pathway, the members of many gene-sets are regulating
%a specific cellular process by interacting in ways which can be summarized by a network. %which can be summarized by a network. Prime
%%examples are signaling pathways. 
These observations motivate a need to 
extend classical GSA in a multivariate direction.

This article is about extending GSA to allow assessment of the 
importance or significance
of gene-sets via
multivariate statistics.
As we describe below, the approach we propose 
 can be interpreted as testing for differences between conditions at the level of networks.
To describe gene-gene networks we use 
graphical models in which the absence/presence of edges corresponds to conditional
independence statements. 
The method we propose for multivariate, network-based gene-set testing is called \emph{NetGSA}.
% and is based on a recently proposed approach for high-dimensional two-sample testing \cite{stadler.diffnet}.
For each gene-set, NetGSA carries out a multivariate, network-based comparison between conditions (details are outlined below), adjusts the obtained p-values for multiple
comparison and finally provides a list of gene-sets ordered by significance. 
%We focus on network comparisons and  initially describe NetGSA without reference to differences in the mean level between conditions.
The results of NetGSA (focusing on differences at the network level) can also be combined with those obtained
from ``classical'' GSA to determine significance in
terms of both networks  \emph{and} change in average gene
expression.

Network-based gene-set analysis is  challenging. It requires  two-sample comparison between networks for typically hundreds of gene-sets and involves issues of high-dimensionality and multiplicity. The intention is not to compare two known networks, but rather to test significance of differences between two estimated networks. High-dimensionality poses severe challenges in this setting since the number of samples is typically small compared to the large
parameter spaces required for describing the networks. This makes estimation of graphical model structure (``network inference") a challenging problem, and the difficulties are inherited in the case of two-sample comparison of estimated networks.

The proposed approach for gene-set testing is based on recently
developed methodology for high-dimensional two-sample testing described in 
\citet{stadler.diffnet} and in particular on an approach called \emph{differential
  network} or \emph{DiffNet} which
   quantifies the difference between two networks inferred from different populations by a p-value. 
   DiffNet is based on the sample splitting technique
introduced by \cite{wasserman2009}: the data is randomly split into two halves,
networks are inferred on one half and significance testing (p-value calculation) is
performed on the other half. This process is repeated many times in
order to prevent a ``p-value lottery'' due
to the arbitrary choice of the data split \citep{meinshausenpval2009}.

To the best of our knowledge, 
DiffNet is currently the only
available approach that allows two-sample testing in high-dimensional
graphical models. Note, that permutation-based tests are computationally not
feasible here as a large number of permutations would be necessary to
compensate the multiple testing correction.
We restrict attention to multivariate comparisons at the gene-set level; that is, we test networks whose nodes are identified with members of gene-sets, and whose edges are within rather than between gene-sets. The general approach of \cite{stadler.diffnet} could in principle be applied to comparison of full, $p$-dimensional distributions or networks between conditions, but this is 
a very difficult high-dimensional problem and is
beyond the scope of this paper.

% from both populations
% into two halves. For both populations networks are inferred on
% the first half of the data. The second half is then used to compare
% these networks using a likelihood-ratio type statistic and p-value can
% be computed based on asymptotic results.
% \begin{itemize}
% \item Split data into two halves. Perform network inference on one
%   half. Compute p-values on second half
% \item Iterate this several time and aggregate resulting p-values
% \end{itemize}

The approach we propose can in principle be used with any graphical model formulation.
However, for NetGSA network analyses
have to be conducted for each gene-set. There are typically hundreds of gene-sets and the
number genes per gene-set can vary from a few up to several dozens of
genes. 
Thus, network inference (NI) approaches used in DiffNet have to be computationally efficient, and chosen and
tuned carefully. %They need to satisfy two technical conditions, the
%\emph{sparsity -} and \emph{screeing property}, in order to render
%significant testing trackable. 
%Besides that they need to be
They need to be able to deal automatically with different numbers of nodes (genes), they
have to satisfy two conditions related to potential overfitting 
(screening and sparsity assumptions). 
We focus on Gaussian graphical models (GGMs) and compare a number of specific approaches for their estimation. 
The overall procedure we propose is computationally efficient and
naively parallelizable: for analysis of a lung cancer gene
  expression dataset reported below, with $d=1208$ genes and $S=216$
  gene-sets, computation required $52$ minutes on a multicore system
  using $50$  cores (each $1.5$ GHz) and $504$ GB shared memory.

Section~\ref{sec:methods} introduces our main methodology for
network-based gene-set testing: Section~\ref{sec:gsa} formulates the
hypothesis testing problem; Section~\ref{sec:diffnet} expands on DiffNet
\citep{stadler.diffnet}; Section~\ref{sec:netgsa} describes our
novel algorithm NetGSA; Section~\ref{sec:networkinference}
discusses different network inference (NI) approaches. In Section~\ref{sec:simulations} we report on simulation results and %The simulation
%section compares the different NI approaches \ref{sec:simulations} and shows the performance of NetGSA. 
 real data examples from cancer biology are investigated in Section~\ref{sec:applications}.

\section{Methods}\label{sec:methods}

\subsection{Notation and set-up}\label{sec:gsa}
Let $\X$ and $\Y$ be matrices of gene expression levels obtained under two conditions.
The dimension of $\X$, $\Y$
are $n_x\times
d$ and $n_y\times d$ respectively, where $d$ denotes the total number of genes under study and $n_x$ and $n_y$ are the
condition-specific sample sizes (throughout we use $x$ and $y$ to denote the two conditions).
Gene-sets are denoted $A_s \subset
\{1,\ldots,d\}$, $s=1,\ldots,S$; gene-sets need not  be disjoint. We denote the probability density function of genes belonging to set
$A_s$ by $f^x_{s}$ and $f^y_{s}$ for conditions $x$ and $y$
respectively; these densities are joint over all genes belonging to
the gene-set and accordingly have dimension $d_s$, where $d_s=|A_s|$.
% The hypothesis of interest is:
% $\mathbf{H}_0: \; \mathcal{F}_{1,A_g}=\mathcal{F}_{2,A_g} \qquad
% \textrm{for}\; g=1,\ldots,G$
% For a specific set $A_s$, our aim is to test whether or not the two
% population arise from the same distribution, in other words, we the
% hypothesis testing problem of interest is
% \begin{eqnarray}
%   \label{eq:nullhypothesis}
%   &\mathbf{H}_0: \; \mathcal{F}^u_{s}=\mathcal{F}^v_{s} \qquad
% \textrm{for}\; s=1,\ldots,S.
% \end{eqnarray}
% Many popular GSA approaches proceed by computing gene-specific two-sample scores
% $z_j$, $j=1,\ldots,k$, aggregating them to
% obtain for all gene-sets a test-statistic $a_s$ and then perform significance
% testing based on these aggregate scores.
% As we already pointed out in the Introduction such approaches are
% limited as they compare the two population only on a univariate
% level. The main focus of this work is to devoted to truly \emph{multivariate} gene-set
% analysis, where differences between two populations are measured in
% terms of gene-gene interaction networks. 

For a specific gene-set $A_s$, our aim is to test whether or not the two
%\marginpar{notation should be more standard}
conditions have different graphical model or network structure. We use Gaussian graphical models (GGM)
to model condition-specific conditional independence structure. 
Vertices in the graphical models for gene-set $A_s$ are identified with members of the gene-set and edges with conditional independence statements between them.
We first present NetGSA focusing on testing network differences only and do not test also for differences in the mean; we 
therefore assume the Gaussian distributions $f^x_s$ and
$f^y_s$ have identical, zero mean but potentially non-identical concentration matrices denoted by $\Omega^x_{s}$ and $\Omega^y_{s}$ respectively. 
The edge structure or network of the corresponding graphical models  are defined by
$$ (\Omega^x_{s})_{j,j'}=0 \Leftrightarrow (j,j') \notin E(G^x_{s})\quad \textrm{and}\quad(\Omega^y_{s})_{j,j'}=0 \Leftrightarrow (j,j') \notin
E(G^y_{s}) $$

where $G^x_{s}$ and $G^y_{s}$ are undirected graphs 
associated with the condition-specific graphical models 
and $E(G)$ denotes the edge set of graph $G$. 

Thus, for  gene-set $A_s$, the NetGSA null hypothesis is
\begin{eqnarray}
  \label{eq:nullhypothesis}
  &\mathbf{H}_{0,s}: \; \Omega^x_{s}=\Omega^y_{s}.
\end{eqnarray}
To test these hypotheses, our strategy is to use \emph{Differential Network} (described in 
Section~\ref{sec:diffnet}) to test each gene-set separately 
and then correct the obtained p-values for multiple
testing. Note that gene-sets contain different numbers of genes
which can vary from moderate to very large. As the sample sizes $n_x$
and $n_y$ are
typically small, testing (\ref{eq:nullhypothesis}) is challenging
and involves issues of
high-dimensionality.

% We point out that (\ref{eq:nullhypothesis}) does not pick-up changes
% between the two conditions due to difference in average gene
% expression. However, it is easy to combine 

\subsection{Differential networks}\label{sec:diffnet}
We recently developed a novel and very general approach for 
high-dimensional two-sample testing \citep{stadler.diffnet}. 
We also
outlined how to use the approach for two-sample comparison of 
graphical models (\emph{differential network} or \emph{DiffNet}). In this Section
we review DiffNet
with reference to the NetGSA context; we refer the reader to \cite{stadler.diffnet} for further 
technical details.

%\subsubsection{Sample Splitting, Network Screening and P-Values}\label{sec:diffnet2}
\subsubsection{Network testing using sample splitting}
Consider a gene-set $A_s$ and corresponding gene expression matrices $\X_{s}$ and $\Y_{s}$ for the two conditions
(of size $n_x \times d_s$ and $n_y \times d_s$ respectively). For each condition, we randomly split the data into two halves $\X_s=(\X^{\rm in}_{s},\X^{\rm out}_{s})$ and $\Y_s=(\Y^{\rm in}_{s},\Y^{\rm out}_{s})$. To test the hypothesis $\mathbf{H}_{0,s}: \Omega^x_{s}=\Omega^y_{s}$, DiffNet proceed in two steps:
\begin{enumerate}
\item Network screening step. Based on the \emph{first half} of the data,
  $\X^{\rm in}_{s}$ and $\Y^{\rm in}_{s}$, networks $\hat G^x_{s}$ and
  $\hat G^y_{s}$ are estimated for each condition separately. In
  addition, a third network $\hat G^{xy}_{s}$ is built using pooled
  data ($\X^{\rm in}_{s}$,$\Y^{\rm in}_{s}$). The latter network
  should provide a good model in the null case of
  no difference between the two conditions. Under the alternative,
  however, modeling both conditions with different graphs $\hat
  G^x_{s}$ and $\hat G^y_{s}$ is beneficial. We propose and compare several ways of inferring the networks $\hat G^x_{s}$, $\hat G^y_{s}$ and $\hat G^{xy}_{s}$ in Section~\ref{sec:networkinference}.
\item P-value calculation step. The networks $\hat G^x_{s}$ and $\hat
  G^y_{s}$ model each condition individually and give rise to a
  log-likelihood $L^{\rm ind}_s=L_{\hat G^x_{s}}+L_{\hat G^y_{s}}$. On
  the other hand, $\hat G^{xy}_{s}$ models pooled data jointly with
  log-likelihood $L^{\rm joint}_s=L_{\hat G^{xy}_{s}}$. We compare the
  individual with the joint model using the score 
\begin{eqnarray}
{\Delta}\mathrm{AIC}_s&=&\mathrm{AIC}^{\rm ind}_s-\mathrm{AIC}^{\rm joint}_s\\ \nonumber
&=&2\left(L^{\rm ind}_s-L^{\rm joint}_s\right)-2(\mathrm{df}^{\rm ind}_s-\mathrm{df}^{\rm joint}_s),
\end{eqnarray}
with degrees of freedom $\mathrm{df}^{\rm ind}_s=2d_s+|E(\hat G^x_{s})|+|E(\hat G^y_{s})|$ and $\mathrm{df}^{\rm joint}_s=d_s+|E(\hat G^{xy}_{s})|$. This score is based on the Akaike Information Criterion (AIC).
We emphasize that all log-likelihoods appearing in ${\Delta}\mathrm{AIC}_s$ are evaluated using the \emph{second
half} of the data and involve maximum likelihood estimation with
constraints given by the graphs.
%$\mathcal{LR^{\rm out}}$ compares the likelihood when assuming that both populations arise from different networks with the likelihood obtained by considering both populations coming from the same distribution. 
If $\X$ and $\Y$ arise from different networks, then we expect
${\Delta}\mathrm{AIC}_s$ to be larger than zero. In fact, it can be shown that ${\Delta}\mathrm{AIC}_s$ tends to infinity for
large sample sizes. On the other hand, under the null
hypothesis, ${\Delta}\mathrm{AIC}_s$ is
asymptotically distributed as a shifted weighted-sum-of-chi-quares
with distribution function $\Psi(\cdot;\nu,\delta)$, weights
$\nu=(\nu_1,\ldots,\nu_r)$ and shift  $\delta=2(\mathrm{df}^{\rm
  ind}_s-\mathrm{df}^{\rm joint}_s)$. As a consequence a p-value for
the hypothesis $H_{0,s}$ can be obtained by
\begin{eqnarray}
\mathrm{p}^{s}&=&1-\Psi(\Delta\mathrm{AIC}_s;\nu,\delta).
\end{eqnarray}
For all details, in particular on the computation of the weights $\nu$, we refer to \cite{stadler.diffnet}.
\end{enumerate}

\subsubsection{Screening and sparsity assumptions}
There are two key assumptions in the above procedure which are mandatory to obtain correct p-values which control the type-I
error. Both involve the network inference approach used in the
first step of DiffNet. Consider a $n\times d $ data matrix $\X$
generated from a GGM with graph $G$. Then, the estimated graph $\hat G(\X)$
has to satisfy:
\begin{itemize}
\item Screening assumption (\textbf{ScA}). The edge set of the inferred graph contains
  the edge set of the true data generating graph: $E(G)\subseteq E(\hat G).$
\item Sparsity assumption (\textbf{SpA}). The inferred networks are
  sparse, i.e., the number of edges $|E(\hat G)|$ is not too large
  compared to the sample size $n$. 
\end{itemize}
Both assumptions are important. \textbf{ScA} guarantees that the
models involved in the test statistic are correctly specified and that
$\Delta\mathrm{AIC}_s$ has asymptotic null-distribution
$\Psi(\cdot;\nu,\delta)$. \textbf{SpA} is necessary to 
ensure maximum likelihood estimation in the second split is well-behaved and to
render the
asymptotic approximation of the null-distribution accurate. 

\subsubsection{Sparsity index}
To monitor sparsity we use
a \emph{sparsity index} $m(\hat G)$ defined as:
\begin{eqnarray}
  \label{eq:sparsityindex}
m(\hat G) &=&\frac{2|E(\hat G)|}{n\times p}.
\end{eqnarray}
This quantity  has a motivation in terms of linear regression:
estimating
the concentration matrix of a GGM with graph $G$ can be done by regressing each
variable (or node) $j$ against all neighbouring variables
$\mathrm{nb}_j = \{ i : (i,j) \in E(G) \} \subset\{1,\ldots,k\}$. If we assume that neighbouring sets are of approximately the same size ($|\mathrm{nb}_j|\approx
\mathrm{const}$) then the inverse $1/m(G)$ of the sparsity index for graph $G$ 
can be interpreted as the number of
samples per predictor. In linear regression a typical rule of thumb is
$5$ to $10$ samples per predictor to obtain well-behaved parameter estimation. 
Later, we use the sparsity index $m$ to carry out adaptive thresholding to ensure estimation
and p-value calculation in the second split are well-behaved. 

In Section~\ref{sec:networkinference} we discuss different GGM estimation procedures from the literature and discuss their properties in terms of the screening and sparsity assumptions.   

\subsection{The NetGSA algorithm}\label{sec:netgsa}
By applying DiffNet to all gene-sets $A_s$, $s=1,\ldots,S$ we get p-values $p^s$, $s=1,\ldots,S$. To test (\ref{eq:nullhypothesis}) we then adjust these values for multiple comparison and obtain the corrected p-values
$\tilde{\mathrm{p}}^{s}$, $s=1,\ldots,S$. The outcome  depends heavily on the initial data splitting (see Section~\ref{sec:diffnet}). Depending on the random split of the data we can get different results which amounts to a ``p-value lottery" \citep{meinshausenpval2009}. To get stable and reproducible results we therefore repeat the splitting process many times and aggregate the resulting p-values. A simple approach combines the p-values by taking the median \citep{vandewiel2009}. %\cite{meinshausenpval2009} propose a more complicated approach. 
Algorithm~\ref{alg:gsadiffnet} summarizes the overall procedure which we call \emph{NetGSA}. 
% the hypothesis $\Omega^u_{s}=\Omega^v_{s}$. We 
% Our strategy is to use DiffNet (``differential network'', see Section~\ref{sec:diffnet}) to perform for each gene-set a separate test and then correcting the obtained p-values for multiple testing. Note, that gene-sets contain a different numbers of genes which can vary from moderate to very large. Therefore, testing \ref{eq:nullhypothesis} involves issues of both multiplicity and high-dimensionality.
% multi-splitting, multiple comparison, p-value aggregation.

% algorithm

% Note: combining 'DiffNet p-values' and 'high-dim T2 p-values'.
\begin{algorithm}                      % enter the algorithm environment
\vspace{0.1cm}
\textbf{Input} Data $\X$ and $\Y$, gene-sets $A_s$, $s=1,\ldots,S$, number of data splits $B$.\vspace{0.1cm}
\begin{algorithmic}[1]         % give the algorithm a caption
                     % and a label for \ref{} commands later in the document
               % enter the algorithmic environment
\State Randomly split data into two halves
\For{$s=1,\ldots,S$} \vspace{0.1cm}

\State \emph{Network Screening (on 1st half)} \vspace{0.2cm}

Using a network inference procedure, infer networks $\hat G^x_s$, $\hat G^y_s$ and $\hat G^{xy}_s$.\vspace{0.2cm}

\medskip
\State \emph{P-values (on 2nd half) }

\medskip
Evaluate $L^{\rm ind}_s=L_{\hat G^x_{s}}+L_{\hat G^y_{s}}$, $L^{\rm joint}_s=L_{\hat G^{xy}_{s}}$ and $\mathrm{df}^{\rm ind}_s$, $\mathrm{df}^{\rm joint}_s$.\vspace{0.2cm}

Evaluate $\Delta\mathrm{AIC}_s=\mathrm{AIC}^{\rm ind}_s-\mathrm{AIC}^{\rm joint}_s.$\vspace{0.2cm}

Obtain $\mathrm{p}^{s}=1-\Psi(\Delta\mathrm{AIC}_s;\nu,\delta)$,

where $\delta=2(\mathrm{df}^{\rm  ind}_s-\mathrm{df}^{\rm joint}_s)$
and $\nu$ are weights 
estimated following

 \cite{stadler.diffnet}.
\medskip
\EndFor
\vspace{0.1cm}
\State Calculate FDR-corrected p-values: $\tilde{\mathrm{p}}^{1},\ldots,\tilde{\mathrm{p}}^{S}$.
\vspace{0.1cm}
\State Repeat steps 1-6 $B$ times. 

%\State Aggregated p-values and test-statistics (median): $\tilde{\mathrm{p}^{\rm med}}_{s}$ $\mathcal{LR}^{\rm med}_s$

  \medskip
\end{algorithmic}
\textbf{Output} Aggregated quantities (median over $B$ splits):
$\tilde{\mathrm{p}}^{s, \rm med}$ ($s=1,\ldots,S$).\vspace{0.2cm}
\caption{NetGSA: Network-based gene-set analysis}  \label{alg:gsadiffnet} 
\end{algorithm}

We point out that Algorithm~\ref{alg:gsadiffnet} tests only for
differences in terms of networks and does not pick-up changes
between the two conditions due to a difference in mean level.
We propose below a simple procedure for combining the results of NetGSA as described above 
with standard approaches for mean-based GSA to obtain a combined gene-set test that captures differences in networks and the mean.

\subsection{Network Inference (NI) Approaches}\label{sec:networkinference}
Network inference (NI) is an important part of DiffNet. In
Section~\ref{sec:diffnet} we pointed out that NI has to be in line
with \textbf{ScA} and \textbf{SpA} in order to obtain valid
p-values. Besides that, NI is the limiting factor in the overall
computational complexity of Algorithm~\ref{alg:gsadiffnet}. Note that
three networks have to be inferred for each of the $S$ gene-sets in each of the $B$  data splits. Thus, in total there are $3\times S\times B$
networks to be estimated. We now consider different NI approaches. The properties of DiffNet with each of these NI methods are examined in Section~\ref{sec:sim.ni}.
\begin{enumerate}
\item \emph{Graphical Lasso} ({\bf GL}). The Graphical Lasso \citep{yuan05model,friedman2007sic}
%L1-regularized maximum likelihood estimation in the Gaussian model
% is widely used for Graph selection and is well-known under the name Graphical Lasso (or GLasso). It 
 estimates the concentration matrix $\Omega$ by optimizing
$$\hat\Omega^{\lambda}=\argmin-\log|\Omega|+\mathrm{tr}(\mathbf{S}\Omega)+\lambda\|\Omega\|_1.$$
where $\mathbf{S}$ is the sample covariance matrix and $\lambda$ denotes a
penalty parameter. Let $G(\hat\Omega^{\lambda})$  denote the graph structure defined by the zero entries of
$\hat\Omega^{\lambda}$. 

The choice of $\lambda$ is very important, as it
determines the sparsity of the graph: too large $\lambda$ leads to a
very sparse solution which is likely to be in contradiction with
\textbf{ScA}. On the other hand, too little regularization (small
$\lambda$) results in dense networks which conflict with
\textbf{SpA}. 

We set $\lambda$ to a value $\lambda^*$ as follows.
We take a sequence of twenty $\lambda$-values on the log-scale between
$\lambda_{\rm max}=\max_{j>j'}|S_{jj'}|$ and $\lambda_{\rm
  min}$, with $\lambda_{\rm min}=\lambda_{\rm max}/100$ ($p>n$) and $\lambda_{\rm min}=\lambda_{\rm max}/1000$ ($p<n$). Then, $\lambda^*$
is selected by
either 10-fold cross-validation or the Bayesian information criterion
(BIC) with degrees of freedom $\mathbf{df}=\sum_{j\geq j'}\mathbf{1}_{\hat
\Omega^{\lambda}_{jj'}\neq 0}$. Both routes aim to select $\lambda$ in a prediction optimal fashion and typically result in relatively sparse networks which
overestimate the true structure but are likely to satisfy
\textbf{ScA}. 

Additionally, to ensure \textbf{SpA} is satisfied, we evaluate the
sparsity index $m(G(\hat\Omega^{\lambda^*}))$. If inverse
sparsity $1/m$ exceeds a pre-defined threshold $\tau$, we set to zero those entries in $\hat\Omega^{\lambda^*}$ having the smallest corresponding absolute partial correlations 
until $1/m \leq \tau$.
%
%threshold 
%
%In practice, however, sometimes the resulting graphs are
%still too dense relative to the sample size available for the 2nd step of DiffNet, which in turn can inflate the type-I error of the obtained p-value. 
%%The reason is that the sparsity assumption is violated and the asymptotic approximation of the null-distribution is not accurate enough. 
%To overcome this potential issue we propose to additionally threshold the smallest
%entries of $\hat\Omega^{\lambda_{\rm opt}}$ such that the inverse sparsity
%index $\mathcal{T}$ (Equation~\ref{eq:sparsityindex}) does not exceed
%a predefined threshold.
% number of nonzero entries (= sparsity index), $$\mathrm{sparsity\;index}=\mathrm{\Big\{|\hat\Omega_{jj'}^{\lambda_{\rm opt},\textrm{thresh}}|>0\Big\}}/(n\times k),$$
% does not exceed a value $\tau$
In all numerical examples we use $\tau
= 5$ which is a reasonable heuristic in light of the regression
interpretation from Section~\ref{sec:diffnet}. 
% The motivation behind this thresholding rule originates from the connection between estimating the concentration matrix and regressing each
% variabe (or node) against all other variables (nodes). In terms of regression,
% the above thresholding rule states that there should be at least $\tau^{-1}$ times
% more samples than active predictors to get reasonable estimates.
% In
% light of the regression interpretation  the
% value $\tau=5$ is a reasonable heuristic. 

For computations of the Graphical Lasso we use
the \textbf{R}-package \texttt{glasso} \citep{friedman2007sic}. We note that BIC is computationally more efficient than cross-validation.
\item \emph{Meinshausen-B\"uhlmann} ({\bf MB}). The Meinshausen-B\"uhlmann approach estimates a sparse
graphical model by fitting a lasso model to each variable, using the
others as predictors \citep{meinshausen04consistent,tibshirani96regression}. The graph structure is given by the non-zero
entries of the estimated regression coefficients. We take for all
Lasso regressions the same tuning parameter $\lambda$ which we
determine by 10-fold cross-validation on a $\lambda$-grid as described above. The MB approach has similar
properties as the Graphical Lasso. In our examples, however, we found that
the MB approach results in sparser graphs. 

MB is performed using the
function \emph{glasso} (\textbf{R}-package \texttt{glasso}) with the option \texttt{approx=TRUE}.
\item \emph{Shrinkage Estimation} ({\bf Shrink}). We further consider the approach proposed by \cite{schaefer2005}
  based on shrinkage estimation of the covariance matrix and
  subsequently testing for non-zero partial correlation coefficients.  This approach is
  computationally very simple. The shrinkage level can be
  obtained analytically following \citet{ledoit2004}  
  and there is no need for CPU expensive tuning parameter selection, e.g., cross-validation. Testing for non-zero
  partial correlation coefficients involves FDR correction.
  However, it is
  known \citep{kraemer2010} that the inferred networks systematically underestimate the number of edges in the true
  graph. Therefore, despite its computational advantages, it is interesting to investigate whether 
  {\bf Shrink} can be used  in step 1 of
  DiffNet. 

We use the \textbf{R}-package
  \texttt{parcor} \citep{kraemer2010} with the \texttt{default}
  FDR cut-off 0.8 to compute networks based on the shrinkage estimator.
\end{enumerate}

\section{Simulations}\label{sec:simulations}
Ability to compare networks is crucial to NetGSA. We
  therefore begin in Section~\ref{sec:sim.ni} by comparing the different NI methods from Section~\ref{sec:networkinference} 
in terms of their performance in testing network differences.
In the following Section~\ref{sec:sim.netgsa} we consider gene-set simulations to test NetGSA itself.

\subsection{Comparison of different network inference approaches}\label{sec:sim.ni}
%We emphasized in Section~\ref{sec:diffnet} that the validity of the
%obtained p-values depends on two important conditions on the network
%inference approach used in the first step of DiffNet. The
%\emph{screening} assumption requires that the estimated graph contains
%all true edges, while the \emph{sparsity} assumption assumes that the
%inferred network is ``sparse enough''.

% Ability to compare networks is crucial to NetGSA. We therefore begin by comparing the different NI methods from Section~\ref{sec:networkinference} 
% in terms of their performance in testing network differences. 
% In the following Section we consider gene-set simulations to test NetGSA itself.

\begin{figure}[htbp!]%--- Bild 'H'ier, 'B'ottom oder 'T'op  ``!'' ich WILL!!
\begin{centering}
   \includegraphics[scale=0.6]{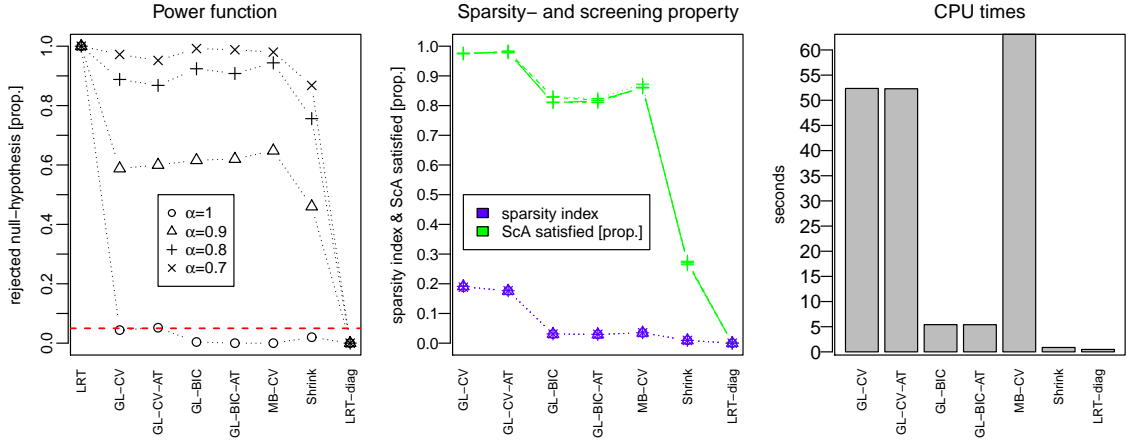} 
  \caption[]%<<-- Legende für 'List of Figures'  (ohne ``[..]'': 2x dasselbe).
  {Results Model 1. Performance of DiffNet with different NI approaches and different levels of network concordance ($\alpha$). First panel: Power function (proportion of rejected
    null-hypothesis). Second panel: In green: proportion of times \textbf{ScA} is
    satisfied, in blue: sparsity index. Third panel: CPU
    times. [The different NI methods \textbf{GL-CV}, \textbf{GL-CV-AT}, \textbf{GL-BIC},
    \textbf{GL-BIC-AT}, \textbf{MB-CV} and \textbf{Shrink} are described
    in Table~\ref{tab:sim1}. \textbf{LRT}: two-sample
    likelihood-ratio test for difference between covariance matrices (asymptotic $\chi^2_{d\times (d+1)/2}$ null-distribution). \textbf{LRT-diag}: diagonal-restricted
    two-sample likelihood-ratio test (asymptotic $\chi^2_d$ distribution)].}% eigentliche Bild-Legende.
\label{fig:mod1}
\end{centering}
\end{figure}

\begin{figure}[htbp!]%--- Bild 'H'ier, 'B'ottom oder 'T'op  ``!'' ich WILL!!
\begin{centering}
   \includegraphics[scale=0.6]{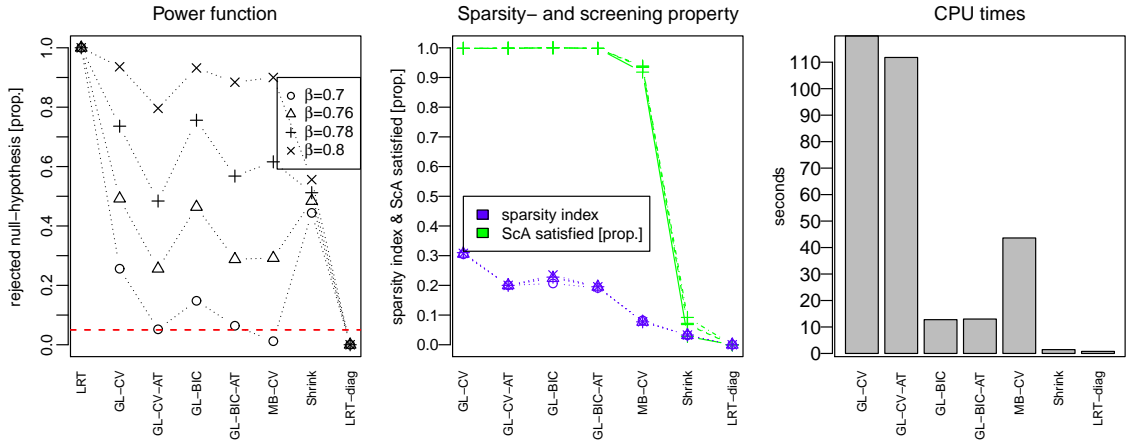} 
  \caption[]%<<-- Legende für 'List of Figures'  (ohne ``[..]'': 2x
            %dasselbe).
{Results Model 2. Same caption as in Figure~\ref{fig:mod1}.}
\label{fig:mod2}
\end{centering}
\end{figure}

To compare NI methods, we generate data
from the following models:

\begin{description}
\item[Model 1] Draw data for the two conditions $x$ and $y$ from a $d$-dimensional multivariate
  normal distribution with $n_x=n_y=100$ and $d=50$. The inverse covariance matrices $\Omega_x$
  and $\Omega_y$ each have $10$ non-zero entries at random locations of which $10\times\alpha$ are at the same
  position in $\Omega_x$
  and $\Omega_y$. We take $\alpha=0.7, 0.8, 0.9$ and $1$,
  where $\alpha=1$ represents the $\mathbf{H}_0$ scenario. Generated
  data are scaled and centered to have zero mean and variance one.

 \item[Model 2] Draw both populations
   according to a $d$-dimensional multivariate normal distribution with
   $n_x=n_y=100$, $d=50$ and 1st order autoregressive covariance matrices. In
   particular, we take $(\Sigma_x)_{jj'}=0.7^{|j-j'|}$ and
   $(\Sigma_y)_{jj'}=\beta^{|j-j'|}$ with $\beta=0.7, 0.76, 0.78,
   0.8$. Then, the choice $\beta=0.7$ corresponds to the
   $\mathbf{H}_0$ assumption. Generated
  data are scaled and centered to have zero mean and variance one.

\end{description}
For each of these models we performed $250$ simulation runs. In each 
run we carried out a DiffNet analysis using the NI approaches
summarized in Table~\ref{tab:sim1}. We report the proportion of rejected
null-hypothesis (power function), the number of times \textbf{ScA} is
satisfied, the relative number of nonzero elements in the inverse
covariance matrices ($\mathrm{sparsity\; index}$, see
equation~(\ref{eq:sparsityindex}) of
Section~\ref{sec:diffnet}) and the CPU time\footnote{This was obtained using the specific {\bf R} packages mentioned above, and is therefore an implementation-dependent measure. We did not consider formal computational complexity.}. We further add as a
reference the performance of two ``classical'' two-sample
likelihood-ratio tests (LRTs) where the first is based on maximum likelihood
estimation with unrestricted covariance matrices (\textbf{LRT}, asymptotic $\chi^2_{d\times(d+1)/2}$ null-distribution) and the second assumes diagonal covariance matrices (\textbf{LRT-diag}, asymptotic $\chi^2_d$ null-distribution). All results are shown in Figures~\ref{fig:mod1} and~\ref{fig:mod2}.

For Model~1, we find that {\bf GL} and {\bf MB} 
approaches control the type-I error at the 5\%-level. They
are also comparable in terms of power. As expected,
selecting $\lambda^{*}$ with BIC is about 10 times
faster then cross-validation. Interestingly, the shrinkage
approach \textbf{Shrink} has type-I error control despite of frequent model
misspecification in step 1 of DiffNet (\textbf{ScA} holds in only 30\% of
the simulation runs satisfied). Nevertheless, \textbf{Shrink} performs
worse in terms of power.

From the results for Model~2 we see that
\textbf{CV} and also \textbf{BIC} show too many false rejections which
can be explained by a larger sparsity index. Use of the adaptive  thresholding procedure we propose above 
(\textbf{GL-CV-AT} and \textbf{GL-BIC-AT}) is sufficient to settle the
false positive rate at the desired 5\%-level. In
Model~2, \textbf{Shrink} performs badly: in about half
of the cases    the null-hypothesis is wrongly rejected. Note, that the screening
assumption is satisfied in less than 10\% of the cases when using
\textbf{Shrink}. This suggests that it may not be advisable
to use \textbf{Shrink} for network inference in DiffNet. Finally, our reference
methods, \textbf{LRT} and \textbf{LRT-diag}, behave as expected: as a
consequence of small sample sizes the asymptotic $\chi^2_{d\times (d+1)/2}$
null-distribution is a very poor approximation when computing
p-values using \textbf{LRT}. On the other hand the likelihood-ratio statistic used in
\textbf{LRT-diag} contains only information about potential differences in
the diagonal of the
covariance matrices and therefore cannot detect any differences
between the two conditions in the examples here. 

From the analysis of the results for both Models~1 and 2 we conclude that \textbf{GL-BIC-AT} is a good
choice and we use it as the default NI
approach in
NetGSA for all our
subsequent analyses.
\begin{table}[h!]
\centering
%\tabcolsep=1.2pt
%  \scriptsize
  \begin{tabular}{c|c|c|c}
Name  & Network inference method &Tuning
    parameter&Adaptive thresholding\\\hline
\textbf{GL-CV}&Graphical Lasso&cross-validation&no\\
    \textbf{GL-CV-AT}&Graphical Lasso&cross-validation&yes\\
 \textbf{GL-BIC}&Graphical Lasso&
    BIC & no\\
\textbf{GL-BIC-AT}&Graphical Lasso&
    BIC & yes\\
\textbf{MB-CV}&Meinshausen-B\"uhlmann& cross-validation& no\\
\textbf{Shrink}&Shrinkage covariance estimator& set analytically & no
 \end{tabular}
  \caption{NI (network inference) approaches compared in simulation
    study of Section~\ref{sec:sim.ni}. 
    Key: {\bf GL}, Graphical Lasso; {\bf MB}, Meinshausen-B\"ulmann
    approach; {\bf CV},    tuning parameter chosen by cross-validation; 
    {\bf BIC},    tuning parameter chosen by BIC; 
    {\bf AT}, use of adaptive thresholding (see text); 
    {\bf Shrink},     shrinkage approach proposed by \cite{schaefer2005} with default FDR cut-off $0.8$.
%    \textbf{cv}: Graphical Lasso estimator,
%    $\lambda$ chosen by cross-validation. \textbf{bic}: Graphical Lasso estimator,
%    $\lambda$ chosen by BIC. \textbf{cv.tau5} and \textbf{bic.tau5}:
%    Like \textbf{cv} and \textbf{bic} but with additional
%    thresholding. \textbf{mb}: Meinshausen B\"ulmann
%    approach, $\lambda$ obtained by cross-validation. \textbf{shrink}:
%    Shrinkage approach proposed by \cite{schaefer2005} with default FDR cut-off $0.8$.
}
\label{tab:sim1}
\end{table}

% \begin{table}[h!]
% \centering
% %\tabcolsep=1.2pt
% %  \scriptsize
%   \begin{tabular}{c|c|c|c}
%     Method &Reg. covariance estimation&Tuning parameter&Thresholding\\\hline
%     \textrm{cv}&Graphical Lasso&cross-validation&none\\
%  \textrm{cv.tau10}&Graphical Lasso&cross-validation& ``$\tau=10$''-rule \\
% \textrm{bic}&Graphical Lasso & BIC & none\\
%  \textrm{bic.tau10}&Graphical Lasso&
%     BIC & ``$\tau=10$''-rule\\
% \textrm{mb}&Meinshausen B\"uhlmann& cross-validation& none\\
% \textrm{shrink}&Shrinkage& analytic formula & fdr-correction
%  \end{tabular}
%   \caption{NI (network inference) approaches compared in simulation
%     study of Section~\ref{sec:sim.ni}. []}
% \label{tab:sim1}
% \end{table}

% \begin{figure}[htbp!]%--- Bild 'H'ier, 'B'ottom oder 'T'op  ``!'' ich WILL!!
% \begin{centering}
%    \includegraphics[scale=0.65]{ar1_scale.pdf} 
%   \caption[]%<<-- Legende für 'List of Figures'  (ohne ``[..]'': 2x dasselbe).
%   {Simulation study, Model 2. Size of p-values. Number of times the
%     screening property is satisfied (frequency). Number of edges of
%     infered graphs. CPU times.[].}% eigentliche Bild-Legende.
% \label{fig:mod2}
% \end{centering}
% \end{figure}
\subsection{Performance of NetGSA}\label{sec:sim.netgsa}
In this study we simulate data for $S=20$ gene-sets. For each gene-set
$s$ we generate data matrices $\X_s$ and $\Y_s$ from $\calN(\mu^x_s,\Sigma^x_s)$ and $\calN(\mu^y_s,\Sigma^y_s)$ with $n=n_x=n_y=40$. The number of genes $d_s$ of gene-set $s$ is  drawn uniformly from $\{20,\ldots,n-1\}$ ($d_s < n$  ensures that the conventional LRT  is well-defined).
The gene-set specific means, $\mu^x_s$ and $\mu^y_s$, are taken to be zero except for the first three gene-sets:
\begin{itemize}
\item Gene-set 1: $\mu^x_1=0$. $\mu^y_{1j}=0.2$, $j=1,\ldots,d_1$.
\item Gene-set 2: $\mu^x_2=0$. $\mu^y_{2j}=0,\; j=1,\ldots, \lceil
  d_2/2\rceil$ and $\mu^y_{2j}=0.4,\; j=1,\ldots, \lfloor d_2/2\rfloor$.
\item Gene-set 3: $\mu^x_3=0$. $\mu^y_{3j}=-0.2,\; j=1,\ldots, \lceil
  d_3/2\rceil$ and $\mu^y_{3j}=0.2,\; j=1,\ldots, \lfloor d_3/2\rfloor$.
\item Gene-sets 4-20: $\mu^x_s=\mu^y_s=0$.
\end{itemize}
All gene-sets have inverse covariance matrices generated as in Model
1. We set $\Omega^x_s$ and $\Omega^y_s$ to have $\lceil d_s/2 \rceil$ non-zero
entries at random locations, of which $\alpha_s\times\lceil d_s/2 \rceil$ are common
across both conditions. The parameter $\alpha_s$ controls relative network concordance and is
chosen as:
\begin{itemize}
\item Gene-sets 1-3: $\alpha=\alpha_1=\alpha_2=\alpha_3 \in \{0, 0.25, 0.5, 1\}.$
\item Gene-sets 4-20: $\alpha_s=1$, i.e. $\Omega^x_s=\Omega^y_s$.
\end{itemize}
Gene-sets one to three exhibit a  difference between conditions in terms of their means. All gene-sets have non-trivial gene networks.
By decreasing the parameter $\alpha$ from one to
zero we introduce additional
network difference for the first three gene-sets . 

The aim of this simulation is to investigate performance of NetGSA in
detecting gene-sets with network difference between $x$
and $y$. We further want to investigate whether  NetGSA can improve 
overall performance in the setting in which conditions differ with respect to both means and networks. For that purpose we
compute p-values using NetGSA; \textbf{Net(SS)} stands for the
single-split approach ($B=1$), \textbf{Net(MS)} denotes the multi-split version ($B=50$). We further run a ``classical'' gene-set analysis using
the approach described in \cite{irizarry2009} (\textbf{Classic})
and combine NetGSA with this approach by reporting the
  minimum of the two p-values
(\textbf{Classic+Net(MS)}). Finally, we compute for each gene-set p-values with the
conventional LRT and correct them for multiple
comparison (\textbf{LRT}).

\begin{figure}[htbp!]%--- Bild 'H'ier, 'B'ottom oder 'T'op  ``!'' ich WILL!!
\begin{centering}
   \includegraphics[scale=0.8]{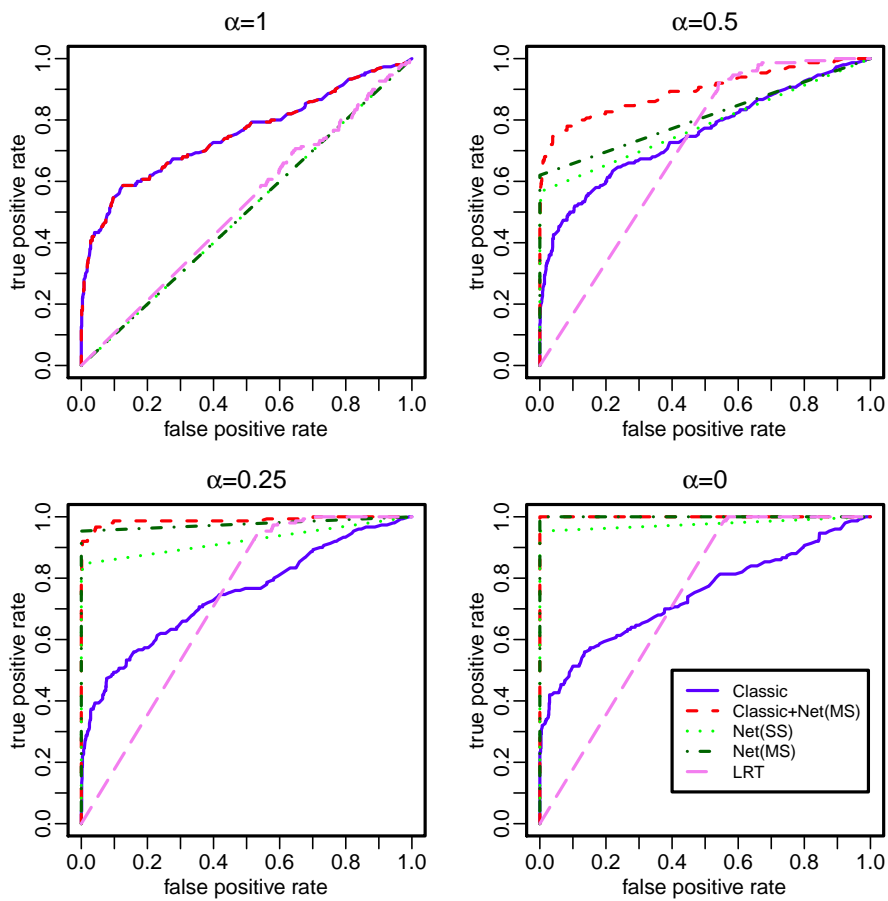} 
  \caption[]%<<-- Legende für 'List of Figures'  (ohne ``[..]'': 2x dasselbe).
  {Simulated data, ROC curves. ROC curves show the average true
    positive rate against the average false positive rate for varying
    thresholds, where the rates are with respect to gene-sets. 
    Each panel corresponds to a different level of overlap between condition-specific networks,
    as controlled by a simulation parameter $\alpha$.
    Upper left panel ($\alpha=1$): no network
    difference. Upper right panel ($\alpha=0.5$): the two networks for
    gene-sets 1-3 share half of the edges. Lower left panel
    ($\alpha=0.25$): the two networks for gene-sets 1-3 share a
    quarter of the edges. Lower right panel ($\alpha=0$): the two networks for gene-sets 1-3
  have no edges in common. [\textbf{Net(SS)}: NetGSA with $B=1$
  (single-split); \textbf{Net(MS)}: NetGSA with $B=50$ (multi-split);
  \textbf{Classic}: classic gene-set analysis as
  described in \cite{irizarry2009};
  \textbf{Classic+Net(MS)}: classic gene-set analysis and NetGSA
  combined; \textbf{LRT}: conventional likelihood-ratio test.]}% eigentliche Bild-Legende.
\label{fig:sim2.exp2}
\end{centering}
\end{figure}

Figure~\ref{fig:sim2.exp2} shows ROC curves, averaged over $50$
simulation runs, for various levels of relative network concordance
$\alpha$. Table~\ref{tab:sim2} shows averaged false discovery and true positive
rates at the $5\%$ significance level. As expected performance of NetGSA improves with increasing
network difference (smaller $\alpha$ values). We also see that
multi-splitting dominates performance compared to using only a
single data split. Testing network differences with \textbf{LRT}
performs poorly in all scenarios. 

\begin{table}[h!]
 \centering
\small
\tabcolsep=2pt
%%  \scriptsize
   \begin{tabular}{|c|c|c|c|c|}\hline
  & \multicolumn{4}{c|}{False discovery rate (false positive rate)}\\\hline
&$\alpha=0$&$\alpha=0.25$&$\alpha=0.5$&$\alpha=1$\\ \hline
\textbf{Classic}&0 (0)&0 (0)&0.02 (0.001)&0 (0)\\\hline
\textbf{Classic+Net(MS)}&0 (0)&0 (0)&0.005 (0.001)&0 (0)\\\hline
\textbf{Net(SS)}&0 (0)&0 (0)&0 (0)&0 (0)\\\hline
\textbf{Net(MS)}&0 (0)&0 (0)&0 (0)&0 (0)\\\hline
\textbf{LRT}&0.851 (0.992)&0.849 (0.994)&0.848 (0.986)&0.848 (0.987)\\\hline
\end{tabular}

\vspace{0.5cm}

\begin{tabular}{|c|c|c|c|c|}\hline
& \multicolumn{4}{c|}{True positive rate}\\\hline
&$\alpha=0$&$\alpha=0.25$&$\alpha=0.5$&$\alpha=1$\\ \hline
\textbf{Classic}&0.08&0.113&0.087&0.107\\\hline
\textbf{Classic+Net(MS)}&0.08&0.447&0.867&1\\\hline
\textbf{Net(SS)}&0&0.387&0.74&0.953\\\hline
\textbf{Net(MS)}&0&0.373&0.86&1\\\hline
\textbf{LRT}&0.987&1&1&1\\\hline
\end{tabular}

 \caption[]%<<-- Legende für 'List of Figures'  (ohne ``[..]'': 2x dasselbe).
  {Performance of NetGSA at the $5\%$ significance level. Average
    false discovery rate (in brackets: false positive rate) and true positive rate at the $5\%$ significance
    level for various methods and different $\alpha$-values (relative
    network concordance of gene-set one to three). [\textbf{Net(SS)}: NetGSA with $B=1$
  (single-split); \textbf{Net(MS)}: NetGSA with $B=50$ (multi-split);
  \textbf{Classic}: classic gene-set analysis as
  described in \cite{irizarry2009};
  \textbf{Classic+Net(MS)}: classic gene-set analysis and NetGSA
  combined; \textbf{LRT}: conventional likelihood-ratio test.]}% eigentliche Bild-Legende.
\label{tab:sim2}
\end{table}

\begin{figure}[htbp!]%--- Bild 'H'ier, 'B'ottom oder 'T'op  ``!'' ich WILL!!
\begin{centering}
   \includegraphics[scale=0.75]{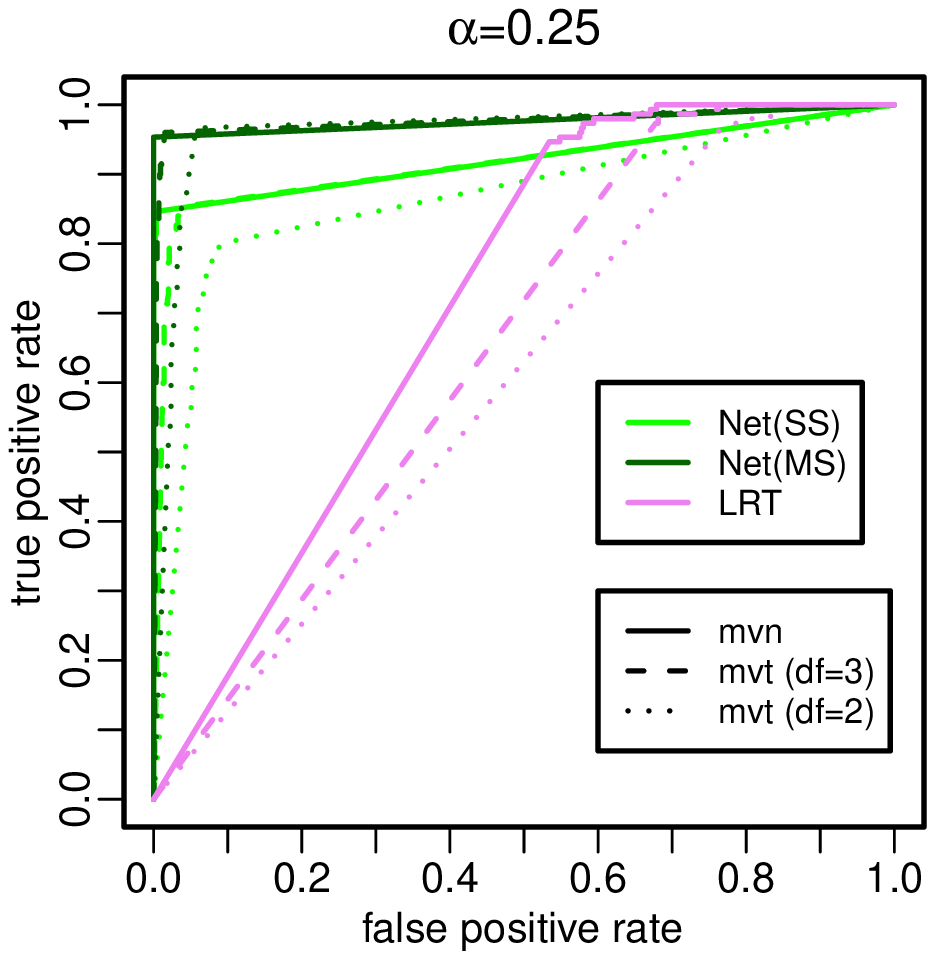} 
  \caption[]%<<-- Legende für 'List of Figures'  (ohne ``[..]'': 2x dasselbe).
  {Effect of t-contamination on performance of NetGSA.  ROC curves plot the average true
positive rate against the average false positive rate for varying
thresholds. Solid lines: multivariate normal data (no contamination). Dashed lines: t-contaminated data
with degree of freedom df=3. Dotted lines: performance for t-contaminated data
with degree of freedom df=2. [\textbf{Net(SS)}: NetGSA with $B=1$
  (single-split); \textbf{Net(MS)}: NetGSA with $B=50$ (multi-split); \textbf{LRT}: conventional likelihood-ratio test.]}% eigentliche Bild-Legende.
\label{fig:sim2.exp1}
\end{centering}
\end{figure}

Calculations in NetGSA are based on
Gaussian graphical models and therefore rely upon the normality assumption. In order to investigate performance under deviations from normality we generate data
as described above but
with~$10\%$ of the data contaminated with samples from a multivariate
t-distribution (degrees of freedom two and three). We take the $\alpha=0.25$ setup and perform $50$ simulation runs. Figure~\ref{fig:sim2.exp1} shows ROC performance of
\textbf{Net(SS)}, \textbf{Net(MS)} and \textbf{LRT} for the cases
with and without t-contamination. As expected performance degrades
with more contamination but the effect is not
dramatic and performance of NetGSA degrades gradually. However, we also see from Table~\ref{tab:sim3} that the average
  false discovery rate is larger than expected in presence of t-contamination. Therefore, we recommend to interpret results with care in case of stronger
deviation from normality as this could severely increase the number of
false discoveries.

\begin{table}[h!]
 \centering
\tabcolsep=2pt
%%  \scriptsize
\small
   \begin{tabular}{|c|c|c|c|}\hline
& \multicolumn{3}{c|}{False discovery rate (false positive rate)}\\\hline
&mvn&mvt (df=3)&mvt (df=2)\\\hline
\textbf{Net(SS)}&0 (0)&0.091 (0.02)&0.299 (0.068)\\\hline
\textbf{Net(MS)}&0 (0)&0.036 (0.007)&0.17 (0.04)\\\hline
\textbf{LRT}&0.848 (0.986)&0.85 (0.998)&0.85 (1)\\\hline
\end{tabular}

\vspace{0.5cm}

\begin{tabular}{|c|c|c|c|}\hline
& \multicolumn{3}{c|}{True positive rate}\\\hline
&mvn&mvt (df=3)&mvt (df=2)\\\hline
\textbf{Net(SS)}&0.74&0.76&0.74\\\hline
\textbf{Net(MS)}&0.86&0.873&0.873\\\hline
\textbf{LRT}&1&1&1\\\hline
\end{tabular}

 \caption[]%<<-- Legende für 'List of Figures'  (ohne ``[..]'': 2x dasselbe).
  {Effect of t-contamination on performance of NetGSA at the 5\%
    significance level. Average
    false discovery rate (in brackets: false positive rate) and true positive rate at the $5\%$ significance
    level for different levels of t-contamination. mvn: multivariate
    normal data (no contamination), mvt (df=3): $10\%$ of the data contaminated with samples from a multivariate
t-distribution with degrees of freedom df=3, mvt (df=2): $10\%$ of the data contaminated with samples from a multivariate
t-distribution with degrees of freedom df=2. [\textbf{Net(SS)}: NetGSA with $B=1$
  (single-split); \textbf{Net(MS)}: NetGSA with $B=50$ (multi-split); \textbf{LRT}: conventional likelihood-ratio test.]}% eigentliche Bild-Legende.
\label{tab:sim3}
\end{table}

\section{Applications}\label{sec:applications}
\begin{figure}[b]%--- Bild 'H'ier, 'B'ottom oder 'T'op  ``!'' ich WILL!!
\begin{centering}
   \includegraphics[scale=0.6]{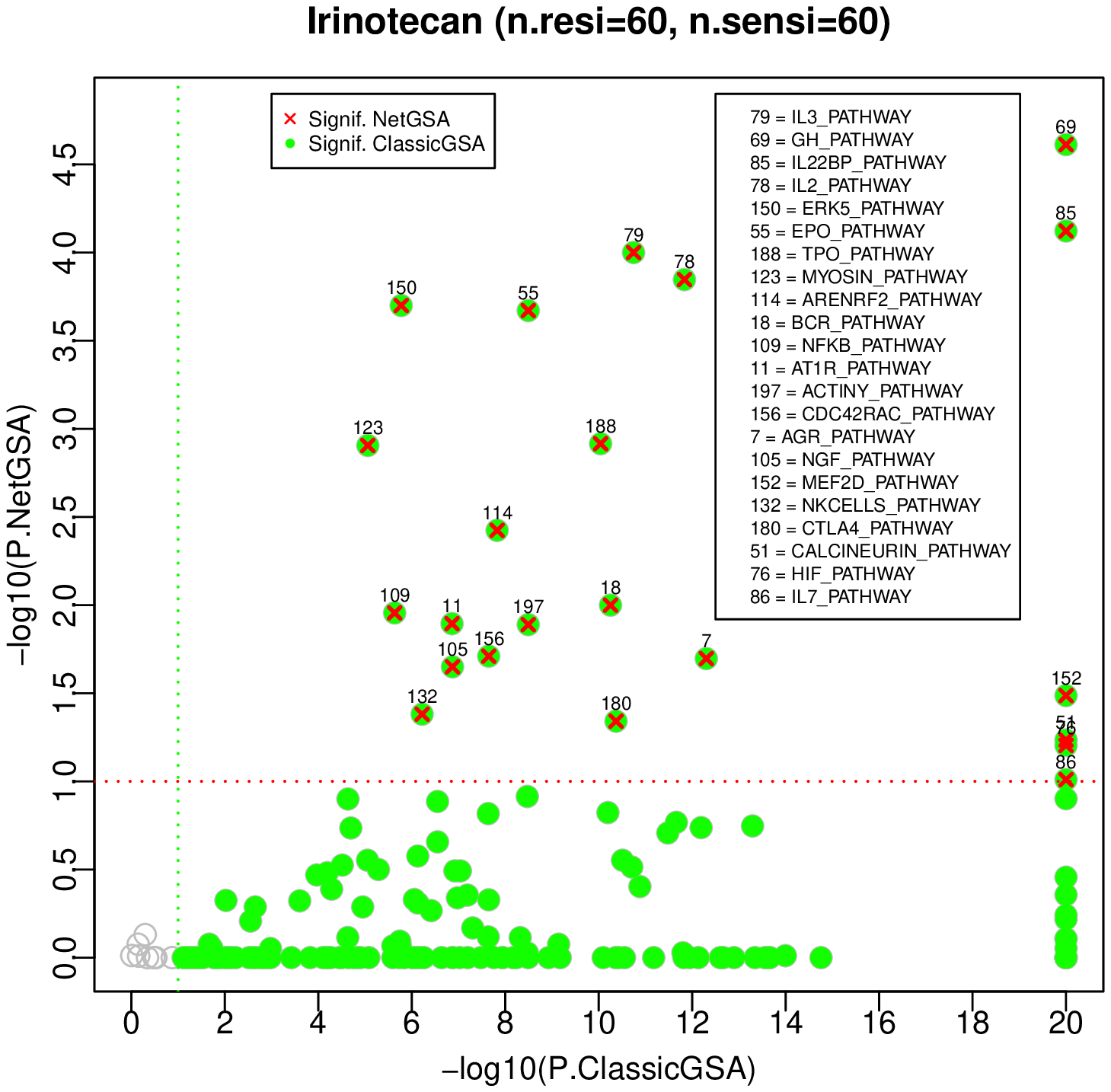} 
  \caption[]%<<-- Legende für 'List of Figures'  (ohne ``[..]'': 2x dasselbe).
  {P-value scatter-plots. CCLE dataset, Irinotecan-resistant
    against -sensitive cell lines. The negative log-p-values obtained
    using NetGSA and classical GSA are plotted against each other. In
    green: significant gene-sets using classical GSA; in red: significant
    gene-sets using NetGSA. At $10\%$ significance level.}% eigentliche Bild-Legende.
\label{fig:scatter_irinotecan}
\end{centering}
\end{figure}

\begin{figure}[b]%--- Bild 'H'ier, 'B'ottom oder 'T'op  ``!'' ich WILL!!
\begin{centering}
   \includegraphics[scale=0.6]{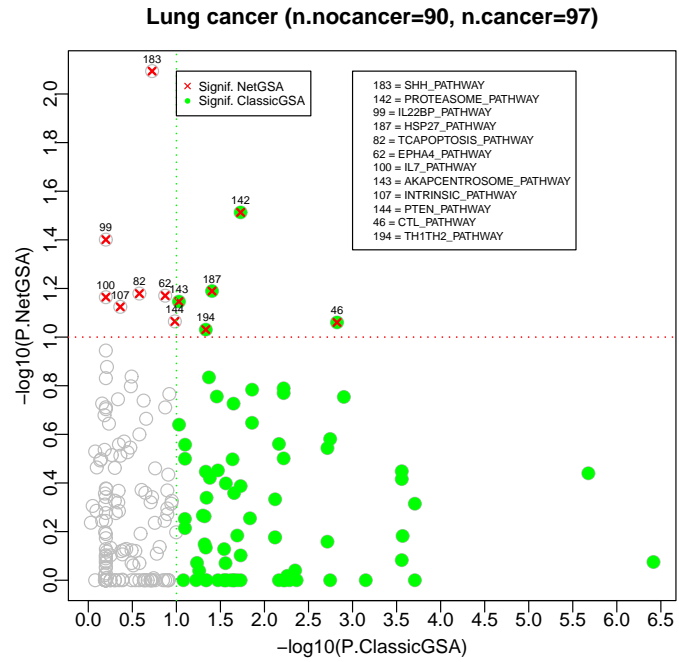} 
  \caption[]%<<-- Legende für 'List of Figures'  (ohne ``[..]'': 2x dasselbe).
  {P-value scatter-plots. Lung cancer dataset, lung cancer subjects
    against control subjects. Same caption as in Figure~\ref{fig:scatter_irinotecan}.}% eigentliche Bild-Legende.
\label{fig:scatter_lung}
\end{centering}
\end{figure}

\begin{figure}[htbp!]%--- Bild 'H'ier, 'B'ottom oder 'T'op  ``!'' ich WILL!!
\begin{centering}
   \includegraphics[scale=0.6]{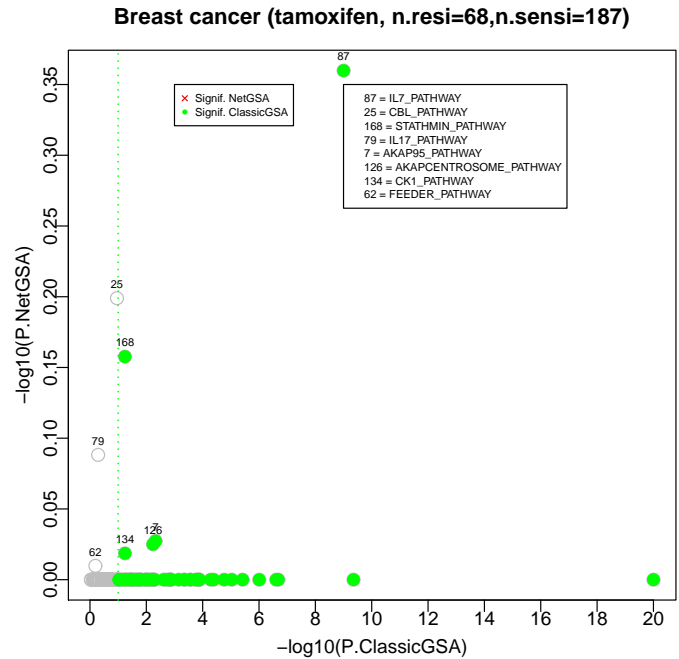} 
  \caption[]%<<-- Legende für 'List of Figures'  (ohne ``[..]'': 2x dasselbe).
  {P-value scatter-plots. Breast cancer dataset, Tamoxifen-resistant
    against -\!~sensitive cell lines. Same caption as in Figure~\ref{fig:scatter_irinotecan}.}% eigentliche Bild-Legende.
\label{fig:scatter_breast}
\end{centering}
\end{figure}

In this Section we apply NetGSA to three datasets from cancer
biology. In each application we compare gene expression data between two conditions, using as gene-sets the collection from \emph{BioCarta} (available at
\url{http://www.broadinstitute.org/gsea/msigdb}). The latter comprises 216 gene-sets with
a number
of genes per gene-set which varies from 6 to 87.
\begin{description}
\item[Cancer cell line encyclopaedia (CCLE)] We consider the dataset
  from the Broad-Novartis Cancer Cell Line Encyclopedia (CCLE) 
  (\cite{barretina2012}\footnote{\url{http://www.broadinstitute.org/ccle}}). Apart from
  gene expression levels the CCLE dataset contains also anticancer drug
  sensitivity measurements. We consider the drug \emph{Irinotecan}. We
  focus on the activity area (AA) and extract the numbers of cell
  lines, $n_{\rm T}$ and $n_{\rm B}$, in the top and bottom third of
  the AA range. We then define 
  resistant and sensitive cell lines by considering the top
  $n_x=\min\{n_{\rm T},n_{\rm B}\}$ scoring cell lines and the bottom
  $n_y=\min\{n_{\rm T},n_{\rm B}\}$ cell lines (with respect to AA). We compare the resistant against the sensitive
  cell lines.

 \item[Lung cancer] We consider gene expression measurements from large
airway epithelial cells sampled from $n_x=97$ patients with lung
cancer and $n_y=90$ controls \citep{spira2007}. This data was previously
analysed with the joint graphical Lasso in \cite{danaher_jgl2013} and
it is publicly available at GEO accession number GSE4115. We compare the
lung cancer samples against the controls.

\item[Breast cancer] The dataset by \cite{loi2008} has
gene expressions from 255 ER+ breast cancer patients
treated with tamoxifen. Using distant metastasis free
survival as a primary endpoint, $n_x=68$ patients from this dataset are labeled
as resistant to tamoxifen and $n_y=187$ are labeled as sensitive to
tamoxifen and we are interested in differences between these two
groups. This dataset is available at GEO accession number GSE6532 and
was analysed using the graph-structured tests for differential
expression proposed by \cite{jacob2012}.

% The first data set by Loi et al. (2008) comprises
% the expression measures of 15,737 genes for 255 ER+ breast cancer patients
% treated with tamoxifen. Breast tumors are generally classified into three
% main categories [Perou et al. (2000)]: luminal epithelial/ER+, HER2+, and
% triple negative. ER+ tumors typically express estrogen receptors at a high
% level and often rely on estrogen for their growth. Tamoxifen is an antagonist
% of estrogen receptors and therefore prevents its activation by endogenous
% estrogen. Some ER+ tumors, however, keep growing after being treated
% with tamoxifen. An important goal is to detect structured groups of genes
% which are differentially expressed between resistant and sensitive patients,
% as detecting such groups could help understand resistance mechanisms and
% eventually improve ER+ breast tumor treatment. Using distant metastasisfree
% survival as a primary endpoint, 68 patients from this data set are labeled
% as resistant to tamoxifen and 187 are labeled as sensitive to tamoxifen.

\end{description}
Our novel
approach, NetGSA, is based on the normality assumption. Strong violation
from normality can result in a inflated false discovery rate. We
therefore conduct Shapiro-Wilk tests for each gene in each condition and discard genes with a p-value (corrected for
multiplicity) smaller than 1\% in either of the two populations. We
run the multi-split version of NetGSA with $B=50$ (\textbf{Net(MS)}) on all three
examples, where we normalized the input
  matrices $\X$ and $\Y$ to have zero mean and variance one. In
  addition to NetGSA we perform classical GSA
(\textbf{Classic}, \cite{irizarry2009}) on unnormalized data. In all analyses we excluded gene-sets with $d_s<5$. Figures~\ref{fig:scatter_irinotecan}-\ref{fig:scatter_breast}
shows scatter plots of the negative log-p-values obtained with \textbf{Net(MS)}
and \textbf{Classic} for the \emph{CCLE}, the \emph{lung
cancer} and the \emph{breast cancer} examples. \textbf{Net(MS)}
identifies fewer significant gene-sets than \textbf{Classic}. Interestingly,
the ordering of gene-sets according to p-values differs substantially
for NetGSA and classical GSA. For example in the lung cancer study there
are several gene-sets significant under \textbf{Net(MS)} but not
under \textbf{Classic}. In order to check that the significant gene-sets are
not false positives we perform ``back-testing'': in particular we pool
data from both populations, randomly divide the data into two
populations and then run NetGSA with only the significant
gene-sets. We repeat this process ten times.  In all examples
``back-testing'' never declares a gene-set as significant.

\begin{figure}[htbp!]%--- Bild 'H'ier, 'B'ottom oder 'T'op  ``!'' ich WILL!!
\begin{centering}
   \includegraphics[scale=0.5]{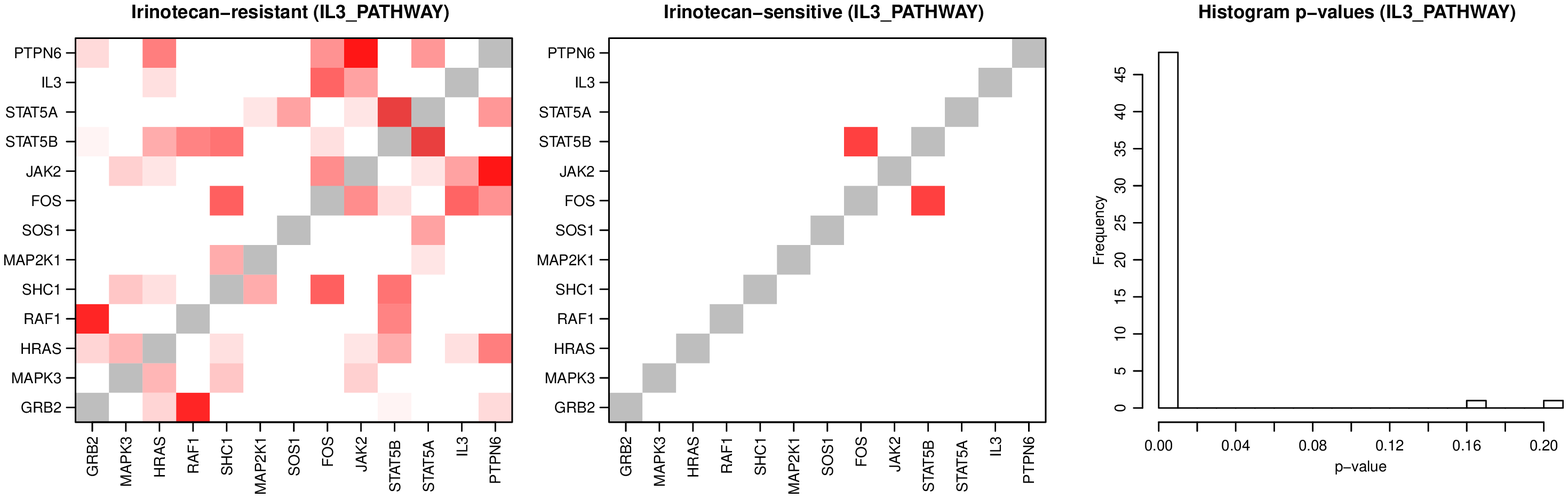} 
  \caption[]%<<-- Legende für 'List of Figures'  (ohne ``[..]'': 2x dasselbe).
  {Networks and distribution of single-split p-values. Left and middle
    panels show heatmaps of the absolute partial correlation
    coefficients (median over $50$ random data splits) of the top gene-set IL3\_PATHWAY for the CCLE
    (Irinotecan) example. The right panel
    shows a histogram of the single-split NetGSA p-values for gene-set IL3\_PATHWAY
    over the $50$ random splits.}% eigentliche Bild-Legende.
\label{fig:il3_pathway}
\end{centering}
\end{figure}

\begin{figure}[htbp!]%--- Bild 'H'ier, 'B'ottom oder 'T'op  ``!'' ich WILL!!
\begin{centering}
   \includegraphics[scale=0.5]{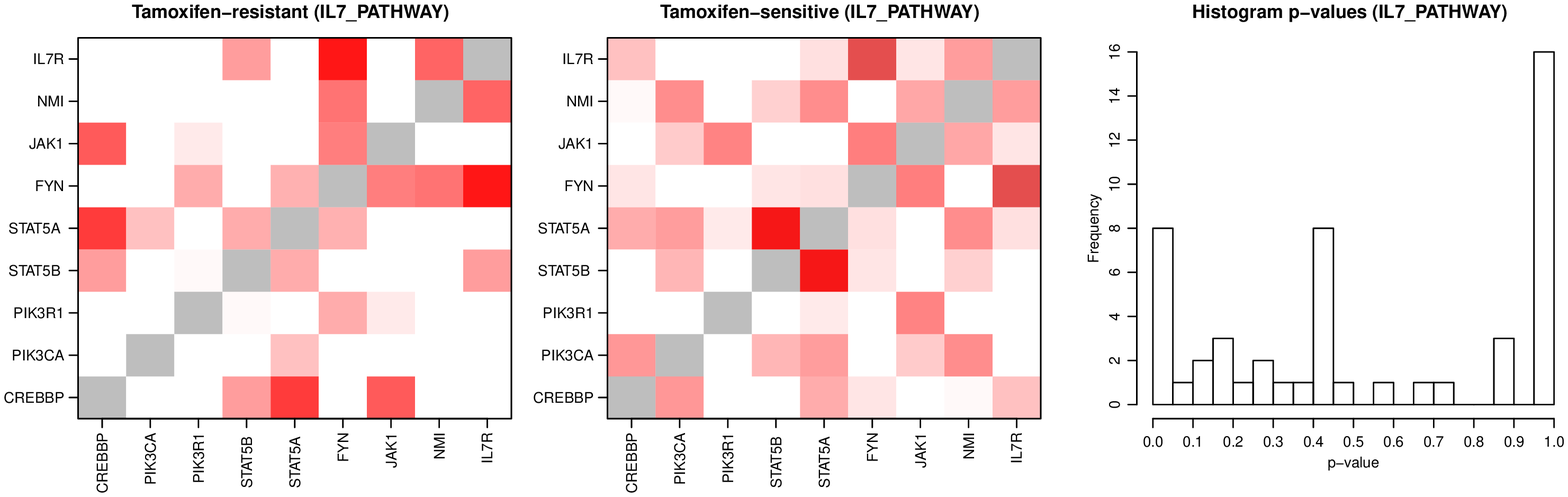} 
  \caption[]%<<-- Legende für 'List of Figures'  (ohne ``[..]'': 2x dasselbe).
  {Networks and distribution of single-split p-values. Left and middle
    panels show heatmaps of the absolute partial correlation
    coefficients (median over $50$ random data splits) of the top gene-set IL7\_PATHWAY for the breast cancer example. The right panel
    shows a histogram of the single-split NetGSA p-values for gene-set IL7\_PATHWAY
    over the $50$ random splits.
}% eigentliche Bild-Legende.
\label{fig:il7_pathway}
\end{centering}
\end{figure}

In Figures~\ref{fig:il3_pathway} and
\ref{fig:il7_pathway} we show the networks (medians of absolute partial
correlation coefficients over 50 random data splits) and a histogram of
single-split p-values over
the fifty random splits for
the top scoring gene-set in each of the three examples. 

\begin{figure}[htbp!]%--- Bild 'H'ier, 'B'ottom oder 'T'op  ``!'' ich WILL!!
\begin{centering}
   \includegraphics[scale=0.5]{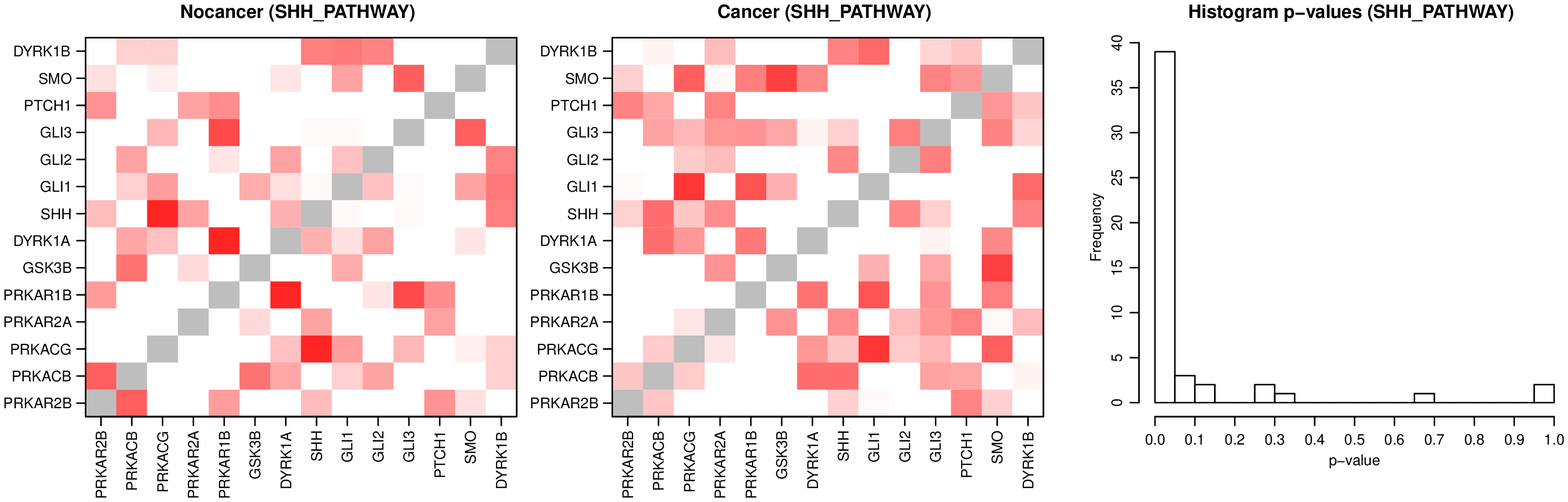} 
  \caption[]%<<-- Legende für 'List of Figures'  (ohne ``[..]'': 2x dasselbe).
  {Networks and distribution of single-split p-values. Left and middle
    panels show heatmaps of the absolute partial correlation
    coefficients (median over $50$ random data splits) of the top gene-set SHH\_PATHWAY for the lung cancer example. The right panel
    shows a histogram of the single-split NetGSA p-values for gene-set SHH\_PATHWAY
    over the $50$ random splits.}% eigentliche Bild-Legende.
\label{fig:shh_pathway}
\end{centering}
\end{figure}

% \begin{figure}[htbp!]%--- Bild 'H'ier, 'B'ottom oder 'T'op  ``!'' ich WILL!!
% \begin{centering}
%    \includegraphics[scale=0.5]{noedges} 
%   \caption[]%<<-- Legende für 'List of Figures'  (ohne ``[..]'': 2x dasselbe).
%   {Number of edges. Left panel: boxplots number of edges for Topotecan
%     resistant and -sensitive populations over significant
%     gene-sets. Middle panel:  boxplots number of edges over nonsignificant
%     gene-sets. Right panel: boxplots number of edges obtained using
%     ``back-testing''  over significant gene-sets.}% eigentliche Bild-Legende.
% \label{fig:noedges}
% \end{centering}
% \end{figure}

\section{Discussion}

The network based gene-set analysis (GSA) procedure we proposed extends gene-set analysis in a multivariate direction. 
We considered a number of network inference (NI) approaches and on the basis of empirical results and computational considerations suggested the use of Graphical Lasso with adaptive thresholding and tuning parameter set using BIC (``GL-BIC-AT"; see Table~\ref{tab:sim1}).

The multivariate nature of our test comes at added computational cost;
our approach is far more computationally demanding that a classical,
mean-based GSA. Nevertheless, 
due to the computational efficiency of Graphical Lasso, simplicity of BIC-based setting of the tuning parameters and parallelizability,
overall the analysis we propose is efficient and practical on a multicore system.
For example, a problem with with $d=1208$ genes and $S=216$
gene-sets took 52 minutes of compute time using 50 cores.
We used GGMs to model and test for differences in network structure. However, 
the high-dimensional two-sample test in \citet{stadler.diffnet} is very general and 
NetGSA could be extended to use other graphical models.
In particular, directed edges can be appropriate in many biological applications, and extension to the case of directed acyclic graphs (DAGs) would be an interesting avenue for future work. Any such extension would have to carefully consider computational demands, since as discussed above NetGSA requires many rounds of network estimation (albeit far fewer than a naive, permutation alternative).

All our examples considered static (i.e. non-time-varying) data. In
many biological applications, time course data play an important
role. NetGSA could be used to compare time-course data between
conditions by use of a suitable graphical model formulation, such as
dynamic Bayesian networks or DBNs \citep{husmeier2003}. In the case of
(a certain class of) DBNs computationally efficient estimation is
possible \citep{hill2012}, and these models could therefore provide a
good starting point for exploring extensions of NetGSA to time-course
data. Such a dynamic variant of NetGSA would allow testing of
differences between gene-sets based on networks estimated from time
course data, and to the extent that time-varying data contain
additional information such an approach could offer improved ability
to detect biologically relevant differences. Naturally, for such an
application sufficient replicates per condition would be needed to
obtain p-values, especially via sample splitting.

In principle, given  sufficiently many samples, network-based testing could be extended beyond the gene-set level to compare networks over all $d$ variables between conditions. In such a set up gene-sets could be used to constrain estimation of the global network, e.g. allowing more edges within gene-sets than between them. Testing could focus on  equality of the overall network, or of subnetworks. NetGSA as proposed here could be regarded as a simplified version of this more general  test, in which only within-gene-set edges are allowed.
%
%However, in practice for gene expression data such an analysis would be challenging due to extreme high-dimensionality. NetGSA could be regarded as 

We considered only gene expression data. However, the methodology could be applied to essentially any molecular data type (including protein and epigenetic), provided suitable sets of variables analogous to gene-sets could be provided. In the case of proteomic data, for example, this could allow testing of differences in protein-protein interplay within known pathways and between conditions.

% time course data
 
 \vspace{1cm}
 
 \noindent
\textbf{Acknowledgements:} The authors would like to thank Rafael Irizarry for discussion concerning classical gene-set testing. This work was supported in part by NCI U54 CA 112970 and the Cancer Systems Biology Center grant from the Netherlands Organisation for Scientific Research.

%The authors thank Peter B\"{u}hlmann for discussions. This work was supported in part by NCI U54 CA 112970 and the
%Cancer Systems Biology Center grant from the Netherlands Organisation for Scientific Research. 

%\bibliographystyle{rss}
%\bibliography{bib_netgsa}

%\appendix
%\begin{figure}[htbp!]%--- Bild 'H'ier, 'B'ottom oder 'T'op  ``!'' ich WILL!!
%\begin{centering}
 %  \includegraphics[scale=0.65]{pscatter_others} 
 % \caption[]%<<-- Legende für 'List of Figures'  (ohne ``[..]'': 2x dasselbe).
  %{P-value scatter-plots. Same caption as in
  %Figure~\ref{fig:topotecan_lung}. Real data:
  %CCLE (Panobinostat, Irinotecan, Paclitaxel) and breast cancer example.}% eigentliche Bild-Legende.
%\label{fig:others}
%\end{centering}
%\end{figure}

\end{document}